\documentclass[superscriptaddress,
  aip,
  pop,
  amsmath,amssymb,
  reprint,
]{revtex4-1}
\usepackage{graphicx,color}
\usepackage{amsmath}
\usepackage{natbib}
\usepackage{epsfig}
\begin{document}

\title{High-frequency elastic moduli of two-dimensional Yukawa fluids and solids}

\author{Sergey Khrapak}
\affiliation{Aix Marseille University, CNRS, Laboratoire PIIM, 13397 Marseille, France}
\affiliation{Institut f\"ur Materialphysik im Weltraum, Deutsches Zentrum f\"ur Luft- und Raumfahrt (DLR), 82234 We{\ss}ling, Germany}
\affiliation{Joint Institute for High Temperatures, Russian Academy of Sciences, 125412 Moscow, Russia}

\author{Boris Klumov}
\affiliation{Aix Marseille University, CNRS, Laboratoire PIIM, 13397 Marseille, France}
\affiliation{Joint Institute for High Temperatures, Russian Academy of Sciences, 125412 Moscow, Russia}
\affiliation{Ural Federal University, 620075 Yekaterinburg, Russia}

\date{\today}

\begin{abstract}
An approach to calculate high-frequency bulk and shear modului of two-dimensional (2D) weakly screened Yukawa fluids and solids is presented. Elastic moduli are directly related to sound velocities and other important characteristics of the system. In this article we discuss these relations and present exemplary calculation of the longitudinal, transverse, and instantaneous sound velocities and derive a differential equation for the Einstein frequency. Simple analytical results presented demonstrate good accuracy when compared with numerical calculations. The obtained results can be particularly useful in the context of 2D colloidal and complex (dusty) plasma monolayers.        
\end{abstract}

\maketitle

\section{Introduction}

In this article we propose a simple and reliable approximation to evaluate high-frequency (instantaneous) elastic moduli of two-dimensional (2D) weakly screened Yukawa fluids and solids. Yukawa systems are of considerable significance in physics, because the Yukawa interaction potential can often be used as a reasonable first approximation to describe actiual interactions between charged particles immersed in a neutralizing medium (e.g. colloidal suspensions, conventional plasmas, and complex or dusty plasmas).~\cite{BelloniJPCM2000,ShuklaRMP2009,IvlevBook,FortovUFN,FortovPR,ChaudhuriSM2011} Elastic moduli are directly related to the sound velocities and
contain important information about the system. For example, measurements of sound velocities in 2D complex plasma (screened Coulomb) crystals have been used to determine the screening parameter and the charge of the particles.~\cite{NunomuraPRL2000,NunomuraPRL2002,NunomuraPRE2002} Many studies have previously addressed important topic related to the thermodynamics and dynamics of 2D Yukawa crystals and fluids.~\cite{PeetersPRA1987,DubinPoP2000,WangPRL2001,TotsujiPRE2004,HartmannPRE2005,VaulinaPRE2010,
SemenovPoP2015,KhrapakPoP08_2015,FengJPD2016,YurchenkoJCP2015,YurchenkoJPCM2016,KryuchkovJCP2017} Most of these studies considered only one phase (either crystal or fluid). The purpose of this work is to present a physically motivated unified approach to evaluate elastic moduli and related properties of 2D Yukawa systems, which represents a good approximation in the strongly coupled fluid regime and becomes exact in the limit of an ideal crystalline lattice. The approach results in simple analytical expressions, very convenient for practical applications. The results can be particularly useful in the context of laboratory experiments with complex plasmas, where the condition of weak screening is usually satisfied.          

\section{Background information}

\subsection{Yukawa systems}

The two-dimensional Yukawa systems considered here are characterized by the repulsive pair-wise interaction potential of the form
\begin{equation}\label{Yukawa}
\phi (r) = (Q^2/r)\exp\left(-r/\lambda\right),
\end{equation}
where $Q$ is the particle charge, $\lambda$ is the screening length,  and $r$ is the separation between two particles. The static properties of Yukawa systems are determined by the two dimensionless parameters: the coupling parameter, $\Gamma = Q^2/a T$, and the screening parameter  $\kappa = a/\lambda$. In the definitions above $T$ is the temperature measured in energy units, $a=(\pi  \rho)^{-1/2}$ is the 2D Wigner-Seitz radius, and $\rho$ is the number density. The coupling parameter is roughly the ratio of the potential energy of interaction between two neighboring particles to their kinetic energy. The system is usually said to be  strongly coupled when this ratio is large, that is $\Gamma\gg 1$. The screening parameter is the ratio of the mean interparticle separation to the screening length. Yukawa systems are considered as weakly screened when $\kappa$ is about unity or below. Strong screening occurs when $\kappa$ is much larger than unity.   

In the strongly coupled regime the system forms a strongly coupled fluid phase, which can crystallize and form a triangular (hexagonal) lattice at a certain $\Gamma=\Gamma_{\rm m}$ (``m'' traditionally refers to melting).   The value of $\Gamma_{\rm m}$ depends on the screening parameter $\kappa$, the approximations for $\Gamma_{\rm m}(\kappa)$ have been proposed in the literature.~\cite{HartmannPRE2005} In the limit $\kappa = 0$ the system reduces to the 2D one-component-plasma (OCP) with $\propto 1/r$ interaction (note that the ``true'' 2D Coulomb interaction, defined as the solution of the 2D Poisson equation, is the logarithmic one~\cite{Caillol1982,KhrapakCPP2016}).  In this case the fluid-solid phase transition occurs at $\Gamma_{\rm m}\simeq 140$.~\cite{GannPRB1979,GrimesPRL1979,KhrapakCPP2016} 

The nature of the fluid-solid phase transition in 2D systems is an important research topic. According to the celebrated  Berezinskii-Kosterlitz-Thouless-Halperin-Nelson-Young (BKTHNY) theory,~\cite{KosterlitzRMP2017} melting in 2D is a two stage process. The crystal first melts by dislocation unbinding to an anisotropic hexatic fluid and then undergoes a continuous transition into the isotropic fluid. This scenario has been confirmed experimentally, in particular for a system with dipole-like, $\phi(r)\propto 1/r^3$, interactions.~\cite{ZahnPRL1999,GrunbergPRL2004,ZanghelliniJPCM2005} One of the most sensitive tests of the BKTHNY theory is the numerical simulations by Kapfer and Krauth~\cite{KapferPRL2015} who studied phase diagram of two-dimensional particle systems interacting with repulsive inverse power-law (IPL, $\propto 1/r^{n}$) and Yukawa interaction potentials. They found that the melting scenario depends critically on the potential softness. For sufficiently soft long-range interactions ($n \lesssim 6$ for IPL and $\kappa\lesssim 6$ for Yukawa) melting proceeds via the BKTHNY scenario. However, for steeper interactions the hard-disk melting scenario with first order hexatic-liquid transition holds.~\cite{BernardPRL2011,EngelPRE2013,ThorneyworkPRL2017} We note that a simplified version the BKTHNY theory has been applied to estimate the location of melting line $\Gamma_{\rm m} (\kappa)$ in Ref.~\onlinecite{PeetersPRA1987}.          

In this study we consider systems weakly screened Yukawa systems with interparticle separations comparable or less than the screening length. To be concrete, we mostly limit consideration to the regime ($\kappa\lesssim 2$). This is the regime particularly relevant to 2D plasma crystals and fluids in laboratory experiments.~\cite{FortovUFN,FortovPR,NunomuraPRE2002,NosenkoPRL2004,NosenkoPRL2009}

\subsection{Elastic moduli}

The high-frequency (instantaneous) elastic moduli of simple (monoatomic) fluids can be expressed in terms of the pair-interaction potential $\phi(r)$ and the radial distribution function (RDF) $g(r)$. Detailed derivation for three-dimensional (3D) fluids was presented by Zwanzig and Mountain.~\cite{ZwanzigJCP1965} The corresponding two-dimensional (2D) analogues are~\cite{IPL3}
\begin{equation}\label{K1}
K_{\infty}=2\rho T-\frac{\pi\rho^2}{4}\int_0^{\infty}drr^2g(r)\left[\phi'(r)-r\phi''(r)\right],
\end{equation}
and
\begin{equation}\label{G1}
G_{\infty}=\rho T+\frac{\pi\rho^2}{8}\int_0^{\infty}drr^2g(r)\left[3\phi'(r)+r\phi''(r)\right],
\end{equation}
where $K_{\infty}$  is the high frequency bulk modulus and $G_{\infty}$ is the high frequency shear modulus. The explicit state-dependence of the RDF, $g(r;\kappa,\Gamma)$ has been omitted for simplicity. Particularly simple derivation for the high-frequency bulk modulus for simple 3D and 2D fluids can be found in Ref.~\onlinecite{KhrapakOnset}. Recently, it has been demonstrated that the expressions relating $K_{\infty}$ and $G_{\infty}$ to $\phi(r)$ and $g(r)$ only work for sufficiently soft interactions and fail when the interaction potential approaches the hard-sphere limit.~\cite{KhrapakSciRep2017} Nevertheless, for the inverse-power-law (IPL) family of potentials $\phi(r)\propto r^{-n}$ this failure occurs only at $n\simeq 20$ (in 3D).~\cite{KhrapakSciRep2017} This implies that for most of actual interactions occurring in nature Eqs. (\ref{K1}) and (\ref{G1}) are still reliable. In particular, they are clearly reliable for weakly screened Yukawa systems studied here.        

The energy $U$ and pressure $P$ (or compressibility $Z$) of 2D monoatomic fluids can be calculated using the integral equations of state~\cite{Hansen_Book,Frenkel2001}
\begin{equation}\label{UPfromG}
\begin{split}
& U= NT\left(1+ \frac{\pi\rho}{T}\int{\phi(r)g(r) r dr}\right),\\
& Z\equiv PV/NT = \left(1 - \frac{\pi \rho}{2T}\int{\phi'(r)g(r) r^2 dr} \right).
\end{split}
\end{equation}

Analyzing the structure of Eqs.~(\ref{K1}) and (\ref{G1}) it is easy to recognize that a certain combination of $K_{\infty}$ and $G_{\infty}$ can be constructed so that $\phi''(r)$ cancels out under the integral. This implies that 
 $K_{\infty}$, $G_{\infty}$  and $P$ are related by the linear equation. Namely,
\begin{equation}\label{CI}
K_{\infty}-2G_{\infty}=2(P-\rho T),
\end{equation}
represents the generalized Cauchy identity for {\it two-dimensional} systems with two-body central interactions, an analogue of the 3D generalized Cauchy identity derived in Ref.~\onlinecite{ZwanzigJCP1965}. This identity applies to arbitrary pair-interaction potentials, provided they are soft enough so that Eqs.~(\ref{K1}) -- (\ref{G1}) are reliable.

\subsection{Sound velocities}

The elastic longitudinal and transverse sound velocities are directly related to the high-frequency elastic moduli. In 2D case we have  
\begin{equation}
m\rho C_{\rm L}^2=K_{\infty}+G_{\infty},  \quad\quad
m\rho C_{\rm T}^2= G_{\infty},
\end{equation}
where $m$ is the atomic (particle) mass and $C_{{\rm L}/{\rm T}}$ is the longitudinal/trensverse sound velocity. The generalized Caushy identity (\ref{CI}) can be expressed in terms of the elastic sound velocities as
\begin{equation}
C_{\rm L}^2-3C_{\rm T}^2=2v_{\rm T}^2(P-\rho T),
\end{equation}
where $v_{\rm T}=\sqrt{T/m}$ is the thermal velocity. Note that expression Eq.~(\ref{CI}) applies also to 3D case, if only potential contribution to the sound velocities are retained~\cite{QCA_Relations} (this is a good approximation for dense fluids and solids). 

Two more quantities to be introduced are the instantaneous (high-frequency)~\cite{Schofield1966} sound velocity related to the instantaneous bulk modulus, 
\begin{equation}
C^2_{\infty}=K_{\infty}/m\rho,
\end{equation}
and the conventional adiabatic sound velocity~\cite{LL_Hydrodynamics} 
\begin{equation}
C^2_{s}=\frac{1}{m}\left(\frac{\partial P}{\partial \rho} \right)_{S}=\frac{K_{\rm S}}{m\rho},
\end{equation}
where $K_{\rm S}$ is the adiabatic bulk modulus. The general inequality $K_{\rm S}\leq K_{\infty}$ was established by Schofield.~\cite{Schofield1966} It has been also demonstrated that $K_{\rm S}$ is in fact extremely close to $K_{\infty}$ for various systems, provided the interaction potential is soft. This includes, for instance, IPL melts in 3D,~\cite{KhrapakSciRep2017} as well as dipole-dipole (IPL3),~\cite{IPL3} Yukawa,~\cite{QCA_Relations} and logarithmic (2D one-component plasma)~\cite{Khrapak2016_2DOCP} interactions in 2D. For weakly screened Yukawa fluids and melts in 3D the longitudinal elastic velocity $C_{\rm L}$ is known to be only slightly higher than the adiabatic sound velocity $C_s$.~\cite{KhrapakPoP08_2015,KhrapakPRE2015_Sound,KhrapakPPCF2016,KhrapakAIPAdv2017}    

\section{Evaluation of elastic moduli in the weakly screened refgime}

Let us now elaborate on the specifics of the Yukawa interaction potential (\ref{Yukawa}). Using the reduced distance $x=r/a$,  introducing the nominal 2D frequency $\omega_0^2=2\pi \rho Q^2/ma$ and recognizing that $\Gamma = \omega_0^2a^2/2v_{\rm T}^2$ we obtain from Eq.~(\ref{UPfromG}) for the excess energy and pressure (contributions associated with the potential interactions)
\begin{equation}\label{excessUP}
\begin{split}
& u_{\rm ex}\equiv U/NT-1=\Gamma\int_0^{\infty}{e^{-\kappa x}g(x)dx},\\
& p_{\rm ex}\equiv Z-1 = \frac{1}{2}\Gamma\int_0^{\infty}{e^{-\kappa x}(1+\kappa x)g(x)dx}.
\end{split}
\end{equation}
In the limit $\kappa \rightarrow 0$ we get $p_{\rm ex}=\tfrac{1}{2}u_{\rm ex}$, as expected for $\propto 1/r$ potential in 2D.

Reduced elastic moduli can be expressed in a similar way, the emerging expressions are       
\begin{equation}\label{Aeq1}
K_{\infty}/\rho T= 2+ \frac{\Gamma}{4}\int_0^{\infty}e^{-\kappa x} (\kappa^2x^2+3\kappa x+3)g(x)dx,
\end{equation}
and
\begin{equation}\label{Aeq2}
G_{\infty}/\rho T= 1+ \frac{\Gamma}{8}\int_0^{\infty}e^{-\kappa x} (\kappa^2x^2-\kappa x-1)g(x)dx.
\end{equation}

\begin{figure}
\includegraphics[width=8cm]{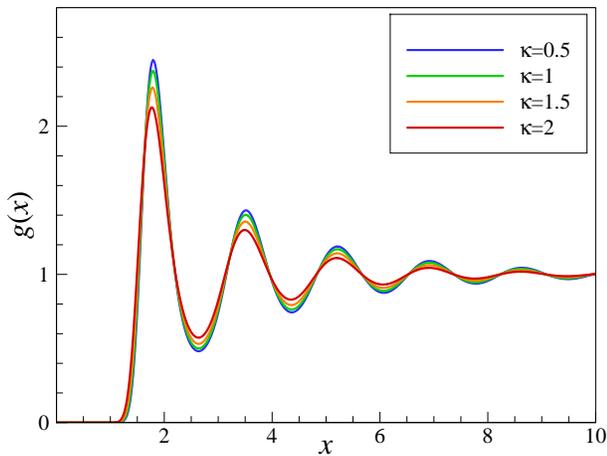}
\caption{Radial distribution functions $g(x)$ corresponding to the strongly coupled 2D Yukawa fluids with $\Gamma = 100$ for several different screening parameters. Some weak dependence on $\kappa$ is present. There are special curves in the phase diagram, usually referred to as isomorphs, along which structure and dynamics in properly reduced units are invariant to a good approximation. These isomorphs are approximately parallel to the melting curve.~\cite{VeldhorstPoP2015} RDFs are quasi-invariant on isomorphs, but not on the lines of constant $\Gamma$. }
\label{Fig1}
\end{figure}

In the regime of weak screening we propose to use the following two simplifications to evaluate elastic moduli, as well as other related thermodynamic quantities of strongly coupled fluids. First, we observe that in the weakly screening regime the RDF $g(r; \kappa, \Gamma)$ is not very sensitive to $\kappa$. For 3D Yukawa systems such an observation was made earlier by Farouki and Hamaguchi~\cite{FaroukiJCP1994} (see, in particular, Figure 6 from their paper). For 2D Yukawa systems this property is illustrated in the Figure~\ref{Fig1}. This property allows us to neglect the dependence of $g(x; \kappa,\Gamma)$ on $\kappa$ when differentiating the excess energy over $\kappa$. This results in the following series of useful relations       
\begin{equation}\label{integrals}
\begin{split}
& \int_0^{\infty}{e^{-\kappa x}g(x)dx} = u_{\rm ex}/\Gamma=f_0(\kappa),\\
& \int_0^{\infty}{\kappa x e^{-\kappa x}g(x)dx= -\left(\kappa/\Gamma\right)\left(\partial u_{\rm ex}/\partial \kappa \right)=f_1(\kappa)}, \\
& \int_0^{\infty}{\kappa^2 x^2 e^{-\kappa x}g(x)dx= \left(\kappa^2/\Gamma\right)\left(\partial^2 u_{\rm ex}/\partial \kappa^2 \right)=f_2(\kappa)}.
\end{split}
\end{equation}

In the equations above we have implicitely used the second simplification. In the weakly screened regime the excess energy only weakly depends on the actual structural properties of the system. This is a general property of soft long-ranged repulsive potentials. Physically, for such potentials the cumulative contribution from large interparticle separations provides dominant contribution to the excess energy. This makes the excess energy insensitive to the actual short-range order. For crystals and fluids not too far from the solid-fluid phase transition the corresponding lattice sum becomes an adequate measure of the excess energy. Mathematically, this can be expressed as
\begin{equation}\label{Madelung}
u_{\rm ex} \simeq M\Gamma,
\end{equation}        
where $M(\kappa)$ is the Madelung coefficient of the triangular lattice. The latter has been evaluated previously.~\cite{TotsujiPRE2004,PereiraPRE2012,KryuchkovJCP2017} The results can be conveniently fitted by the expression~\cite{KryuchkovJCP2017}
\begin{equation}\label{Madelung}
M = -1.1061+0.5038\kappa-0.11053\kappa^2+0.00968\kappa^3+\frac{1}{\kappa}.
\end{equation}  
This fit is constructed in such a way that if the contribution to the excess energy from the neutralizing background is added (which amounts to $-\Gamma/\kappa$) and the limit $\kappa\rightarrow 0$ is considered, the excess energy reduces to that of the 2D OCP with 3D Coulomb ($\propto 1/r$) interaction,~\cite{GannPRB1979,BausPR1980} $u_{\rm ex}\simeq -1.1061\Gamma$. Thus, the $\kappa$-dependent Madelung coefficient fully defines the function $f_0(\kappa)\equiv M$.
Other functions $f_1(\kappa)$, $f_2(\kappa)$, etc. can be trivially obtained when $f_0(\kappa)$ is specified. 

Finally, it is not difficult to recognize that the two simplifications designed to evaluate elastic moduli and related quantities of weakly screened strongly coupled Yukawa fluids become exact in the special case of cold crystalline solid.  In this latter case the RDF $g(x)$ consists of well defined delta-peaks corresponding to the given lattice structure and is thus fixed as long as the lattice is fixed. Mathematically, the function $g(x; \kappa,\Gamma)$ does not anymore depend on $\kappa$ and $\Gamma$ in this limit.     
Moreover, the reduced excess energy is $u_{\rm ex}=M\Gamma+1$ and thus is given exactly by the corresponding lattice sum
in the limit $T\rightarrow 0$ ($\Gamma\rightarrow \infty$). 

Therefore, the method we propose is expected to be a good approximation for strongly coupled fluids and exact in the limit of ideal crystalline lattice. Main results of applying this method to Yukawa fluids and solids are summarized below.    

\section{Results}

\begin{figure}
\includegraphics[width=8cm]{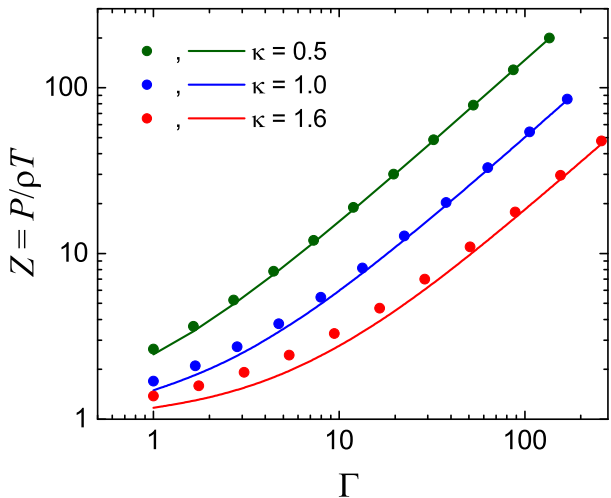}
\caption{Reduced pressure $Z= P/\rho T$ as a function of the coupling parameter $\Gamma$ for 2D Yukawa fluids. Symbols are the results from MD simulations tabulated in Ref.~\onlinecite{KryuchkovJCP2017}. Curves correspond to Eq.~(\ref{pressure}). Three cases correspond to the weakly screening regime with $\kappa = 0.5, 1.0$, and $1.5$. }
\label{Fig2}
\end{figure}

The first obvious emerging expression is that for the excess pressure:
\begin{equation}\label{pressure}
p_{\rm ex} = \frac{\Gamma}{2}\left[f_0(\kappa)+f_1(\kappa)\right].
\end{equation}  
Its accuracy serves as an important check of the reliability of the proposed approximation. Comparison between the calculation using formula (\ref{pressure}) and accurate MD results for the pressure of 2D Yukawa fluids tabulated by Kryuchkov {\it et al.}~\cite{KryuchkovJCP2017} is shown in Fig.~\ref{Fig2}. As expected, the agreement between MD results and Eq.~(\ref{pressure}) is very good at weak screening ($\kappa\lesssim 1$), but worsens as $\kappa$ increases. At a given $\kappa$ the agreement improves as coupling increases. Overall, the provided comparison justifies application of the proposed approximation to dense fluids not too far from the fluid-solid phase transition, provided the screening is sufficiently weak. In the limit $\Gamma\rightarrow \infty$, Eq.~(\ref{pressure}) becomes exact.        

\begin{figure}
\includegraphics[width=8cm]{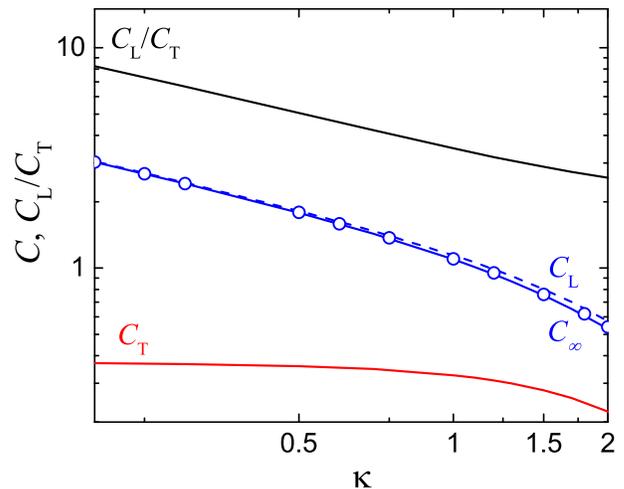}
\caption{Longitudinal ($C_{\rm L}$) and transverse ($C_{\rm T}$) elastic sound velocities of strongly coupled 2D Yukawa systems versus the screening parameter $\kappa$. The instantaneous sound velocity $C_{\infty}$ is very close, but slightly below $C_{\rm L}$ in the considered weakly screened regime. Symbols correspond to the adiabatic sound velocity $C_s$ of 2D Yukawa melts (fluid at $\Gamma=\Gamma_{\rm m}$), calculated using the approach of Ref.~\onlinecite{SemenovPoP2015}. As expected, to a very good accuracy $C_{\infty}$ and $C_s$ coincide. All velocities are expressed in units of $C_0=\sqrt{Q^2/ma}$. The upper curve corresponds to the ratio $C_{\rm L}/C_{\rm T}$. }
\label{Fig3}
\end{figure}

The expressions for the elastic moduli follow directly from Eqs. (\ref{Aeq1}) and (\ref{Aeq2}).  We find it useful to express them in terms of sound velocities. We get
\begin{equation}\label{C1}
C_{\infty}^2/v_{\rm T}^2 = 2+\frac{\Gamma}{4}\left[f_2(\kappa)+3f_1(\kappa)+3f_0(\kappa)\right], 
\end{equation}   
\begin{equation}\label{C1}
C_{\rm L}^2/v_{\rm T}^2 = 3+\frac{\Gamma}{8}\left[3f_2(\kappa)+5f_1(\kappa)+5f_0(\kappa)\right], 
\end{equation}   
and
\begin{equation}\label{C2}
C_{\rm T}^2/v_{\rm T}^2 = 1+\frac{\Gamma}{8}\left[f_2(\kappa)-f_1(\kappa)-f_0(\kappa)\right]. 
\end{equation}   
In the strongly coupled regime the first (kinetic) terms are numerically very small compared to the potential terms and can be neglected. Then, $C_{\infty}$, $C_{\rm L}$, and $C_{\rm T}$ can be expressed as a velocity scale $C_0=\sqrt{Q^2/ma}$ multiplied by certain functions of a single parameter $\kappa$. The ratios of the sound velocities also depend solely on $\kappa$. This serves as a basis behind important methods to estimate particle charge $Q$ and screening parameter $\kappa$ in experiments with complex plasma monolayers.~\cite{NunomuraPRL2002,NunomuraPRE2002} The results obtained within our approximation are plotted in Figure~\ref{Fig3}.    

\begin{figure}
\includegraphics[width=8cm]{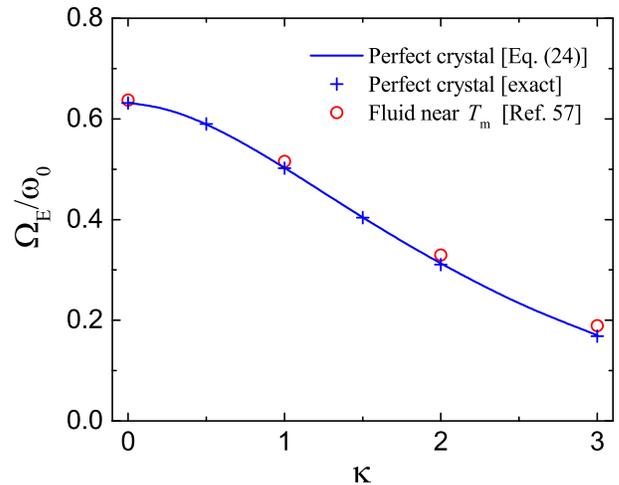}
\caption{Ratio of the Einstein frequency $\Omega_{\rm E}$ to the 2D plasma frequency $\omega_0$ as a function of the screening parameter $\kappa$. The solid curve displays the solution of Eq.~(\ref{Einst_dif}) supplemented by Eq.~(\ref{Madelung}). Crosses are exact results of summation over the perfect hexagonal lattice. Open circles represent numerical results~\cite{DonkoJPCM2008} for strongly coupled Yukawa fluids with $\Gamma_{\rm eff}=120$   (the effective coupling parameter $\Gamma_{\rm eff}=\Gamma f(\kappa)$ was constructed in Ref.~\onlinecite{HartmannPRE2005} by prescribing a constant amplitude for the first peak of the RDF for fixed values of $\Gamma_{\rm eff}$).}
\label{Fig4}
\end{figure}

One more important quantity is the Einstein frequency, $\Omega_{\rm E}$. The Einstein frequency was shown to be an important characteristic of particle dynamics in strongly coupled complex (dusty) plasmas.~\cite{KnapekPRL2007}   Experimental measurements of the ratio $\Omega_{\rm E}/\omega_0$ have been reported recently.~\cite{WongIEEE2017} Here we derive a practical expression for this ratio emerging within the present approximation. The generic expression for the Einstein frequency in 2D is
\begin{equation}
\Omega_{\rm E}^2=\frac{\rho}{2m}\int{d{\bf r}\Delta\phi(r)g({\bf r})}, 
\end{equation}    
where $\Delta \phi (r)=\phi''(r)+\phi'(r)/r$. This expressions applies to both solid and fluid phase, in which case the RDF is isotropic $g({\bf r})=g(r)$. We keep this notation also for the crystalline lattice for simplicity and arrive at
\begin{equation}
\Omega_{\rm E}^2=\frac{\omega_0^2}{2}\int_0^{\infty}\frac{g(x)dx}{x^2}e^{-\kappa x}\left(1+\kappa x +\kappa^2 x^2\right).
\end{equation}   
In the limit $\kappa\rightarrow 0$ we get 
\begin{displaymath}
\Omega_{\rm E}^2=\frac{\omega_0^2}{2}\int_0^{\infty}\frac{g(x)dx}{x^2}.
\end{displaymath}
For an ideal crystalline lattice, the integral $\int_0^{\infty}g(x)dx/x^2$ denotes nothing, but the {\it lattice sum} for the dipole-dipole ($\propto 1/r^{3}$) interaction. For the triangular lattice this sum was evaluated previously,~\cite{IPL3,Topping,derHoff} $M_{\rm IPL3}\simeq 0.798512$.
Hence, in the limit of 2D OCP with the $\propto 1/r$ interaction we get
\begin{equation}
\Omega_{\rm E}^2=0.399256 \omega_0^2.
\end{equation}  
This {\it exact} proportionality coefficient is very close to that of $0.39925$ quoted by Donko {\it et al}.~\cite{DonkoJPCM2008} Differentiating $\Omega_{\rm E}^2$ with respect to $\kappa$ yields
\begin{equation}
\frac{\partial \Omega_{\rm E}^2}{\partial \kappa}=\frac{\omega_0^2}{2}\int_0^{\infty}\frac{g(x)dx}{x^2}e^{-\kappa x}\left(1 - \kappa x\right),
\end{equation}   
which can be further rewritten as
\begin{equation}\label{Einst_dif}
\frac{\partial \Omega_{\rm E}^2}{\partial \kappa}= - \frac{\omega_0^2 \kappa^3}{2\Gamma}\frac{\partial}{\partial \kappa} \left( \frac{u_{\rm ex}}{\kappa}\right).
\end{equation}
This is the exact relation. Approximation (\ref{Madelung}) can be used to derive the explicit analytical expression for $\Omega_{\rm E}$. The result is plotted in Fig.~\ref{Fig4} along with numerical calculations of the lattice sums involved (crosses). Also shown are the results for $\Omega_{\rm E}$ in the strongly coupled fluid regime (circles),~\cite{DonkoJPCM2008} indicating that the Einstein frequency does not change much across the fluid-solid phase transition. This is consistent with a recently reported experimental measurement.~\cite{WongIEEE2017}

\section{Conclusion}

We have proposed a simple and reliable approximation to evaluate high-frequency (instantaneous) bulk and shear moduli of 2D weakly screened Yukawa systems. The approach delivers good accuracy in the strongly coupled fluid regime and becomes essentially exact for an ideal crystalline lattice. As an example of approach application, elastic longitudinal and transverse, as well as the instantaneous sound velocities have been calculated. Similarly to other soft interactions, the longitudinal sound velocity is slightly higher than the instantaneous, while the latter is extremely close to the conventional adiabatic sound velocity.  In addition, we have derived a simple differential equation for the Einstein frequency of 2D Yukawa systems. The solution to this equation reproduces very well the results of numerical calculation of the corresponding lattice sums. The Einstein frequency is only slightly higher for fluid than for an ideal crystal.

\begin{acknowledgments}
This work was supported by the A*MIDEX project (Nr.~ANR-11-IDEX-0001-02) funded by the French Government ``Investissements d'Avenir'' program managed by the French National Research Agency (ANR). We thank V. Nosenko for careful reading of the manuscript. 
\end{acknowledgments}

\bibliographystyle{aipnum4-1}
\bibliography{Moduli_References}

\providecommand{\noopsort}[1]{}\providecommand{\singleletter}[1]{#1}%
\begin{thebibliography}{57}%
\makeatletter
\providecommand \@ifxundefined [1]{%
 \@ifx{#1\undefined}
}%
\providecommand \@ifnum [1]{%
 \ifnum #1\expandafter \@firstoftwo
 \else \expandafter \@secondoftwo
 \fi
}%
\providecommand \@ifx [1]{%
 \ifx #1\expandafter \@firstoftwo
 \else \expandafter \@secondoftwo
 \fi
}%
\providecommand \natexlab [1]{#1}%
\providecommand \enquote  [1]{``#1''}%
\providecommand \bibnamefont  [1]{#1}%
\providecommand \bibfnamefont [1]{#1}%
\providecommand \citenamefont [1]{#1}%
\providecommand \href@noop [0]{\@secondoftwo}%
\providecommand \href [0]{\begingroup \@sanitize@url \@href}%
\providecommand \@href[1]{\@@startlink{#1}\@@href}%
\providecommand \@@href[1]{\endgroup#1\@@endlink}%
\providecommand \@sanitize@url [0]{\catcode `\\12\catcode `\$12\catcode
  `\&12\catcode `\#12\catcode `\^12\catcode `\_12\catcode `\%12\relax}%
\providecommand \@@startlink[1]{}%
\providecommand \@@endlink[0]{}%
\providecommand \url  [0]{\begingroup\@sanitize@url \@url }%
\providecommand \@url [1]{\endgroup\@href {#1}{\urlprefix }}%
\providecommand \urlprefix  [0]{URL }%
\providecommand \Eprint [0]{\href }%
\providecommand \doibase [0]{http://dx.doi.org/}%
\providecommand \selectlanguage [0]{\@gobble}%
\providecommand \bibinfo  [0]{\@secondoftwo}%
\providecommand \bibfield  [0]{\@secondoftwo}%
\providecommand \translation [1]{[#1]}%
\providecommand \BibitemOpen [0]{}%
\providecommand \bibitemStop [0]{}%
\providecommand \bibitemNoStop [0]{.\EOS\space}%
\providecommand \EOS [0]{\spacefactor3000\relax}%
\providecommand \BibitemShut  [1]{\csname bibitem#1\endcsname}%
\let\auto@bib@innerbib\@empty
\bibitem [{\citenamefont {Belloni}(2000)}]{BelloniJPCM2000}%
  \BibitemOpen
  \bibfield  {author} {\bibinfo {author} {\bibfnamefont {L.}~\bibnamefont
  {Belloni}},\ }\href {\doibase 10.1088/0953-8984/12/46/201} {\bibfield
  {journal} {\bibinfo  {journal} {J. Phys.: Condens. Matter}\ }\textbf
  {\bibinfo {volume} {12}},\ \bibinfo {pages} {R549} (\bibinfo {year}
  {2000})}\BibitemShut {NoStop}%
\bibitem [{\citenamefont {Shukla}\ and\ \citenamefont
  {Eliasson}(2009)}]{ShuklaRMP2009}%
  \BibitemOpen
  \bibfield  {author} {\bibinfo {author} {\bibfnamefont {P.~K.}\ \bibnamefont
  {Shukla}}\ and\ \bibinfo {author} {\bibfnamefont {B.}~\bibnamefont
  {Eliasson}},\ }\href {\doibase 10.1103/revmodphys.81.25} {\bibfield
  {journal} {\bibinfo  {journal} {Rev. Mod. Phys.}\ }\textbf {\bibinfo {volume}
  {81}},\ \bibinfo {pages} {25} (\bibinfo {year} {2009})}\BibitemShut {NoStop}%
\bibitem [{\citenamefont {Ivlev}\ \emph {et~al.}(2012)\citenamefont {Ivlev},
  \citenamefont {Lowen}, \citenamefont {Morfill},\ and\ \citenamefont
  {Royall}}]{IvlevBook}%
  \BibitemOpen
  \bibfield  {author} {\bibinfo {author} {\bibfnamefont {A.}~\bibnamefont
  {Ivlev}}, \bibinfo {author} {\bibfnamefont {H.}~\bibnamefont {Lowen}},
  \bibinfo {author} {\bibfnamefont {G.}~\bibnamefont {Morfill}}, \ and\
  \bibinfo {author} {\bibfnamefont {C.~P.}\ \bibnamefont {Royall}},\
  }\href@noop {} {\emph {\bibinfo {title} {Complex Plasmas and Colloidal
  Dispersions: Particle-Resolved Studies of Classical Liquids and Solids}}}\
  (\bibinfo  {publisher} {World Scientific},\ \bibinfo {year}
  {2012})\BibitemShut {NoStop}%
\bibitem [{\citenamefont {Fortov}\ \emph {et~al.}(2004)\citenamefont {Fortov},
  \citenamefont {Khrapak}, \citenamefont {Khrapak}, \citenamefont {Molotkov},\
  and\ \citenamefont {Petrov}}]{FortovUFN}%
  \BibitemOpen
  \bibfield  {author} {\bibinfo {author} {\bibfnamefont {V.~E.}\ \bibnamefont
  {Fortov}}, \bibinfo {author} {\bibfnamefont {A.~G.}\ \bibnamefont {Khrapak}},
  \bibinfo {author} {\bibfnamefont {S.~A.}\ \bibnamefont {Khrapak}}, \bibinfo
  {author} {\bibfnamefont {V.~I.}\ \bibnamefont {Molotkov}}, \ and\ \bibinfo
  {author} {\bibfnamefont {O.~F.}\ \bibnamefont {Petrov}},\ }\href {\doibase
  10.3367/ufnr.0174.200405b.0495} {\bibfield  {journal} {\bibinfo  {journal}
  {Phys.-Usp.}\ }\textbf {\bibinfo {volume} {47}},\ \bibinfo {pages} {447 }
  (\bibinfo {year} {2004})}\BibitemShut {NoStop}%
\bibitem [{\citenamefont {Fortov}\ \emph {et~al.}(2005)\citenamefont {Fortov},
  \citenamefont {Ivlev}, \citenamefont {Khrapak}, \citenamefont {Khrapak},\
  and\ \citenamefont {Morfill}}]{FortovPR}%
  \BibitemOpen
  \bibfield  {author} {\bibinfo {author} {\bibfnamefont {V.~E.}\ \bibnamefont
  {Fortov}}, \bibinfo {author} {\bibfnamefont {A.}~\bibnamefont {Ivlev}},
  \bibinfo {author} {\bibfnamefont {S.}~\bibnamefont {Khrapak}}, \bibinfo
  {author} {\bibfnamefont {A.}~\bibnamefont {Khrapak}}, \ and\ \bibinfo
  {author} {\bibfnamefont {G.}~\bibnamefont {Morfill}},\ }\href@noop {}
  {\bibfield  {journal} {\bibinfo  {journal} {Phys. Rep.}\ }\textbf {\bibinfo
  {volume} {421}},\ \bibinfo {pages} {1} (\bibinfo {year} {2005})}\BibitemShut
  {NoStop}%
\bibitem [{\citenamefont {Chaudhuri}\ \emph {et~al.}(2011)\citenamefont
  {Chaudhuri}, \citenamefont {Ivlev}, \citenamefont {Khrapak}, \citenamefont
  {Thomas},\ and\ \citenamefont {Morfill}}]{ChaudhuriSM2011}%
  \BibitemOpen
  \bibfield  {author} {\bibinfo {author} {\bibfnamefont {M.}~\bibnamefont
  {Chaudhuri}}, \bibinfo {author} {\bibfnamefont {A.~V.}\ \bibnamefont
  {Ivlev}}, \bibinfo {author} {\bibfnamefont {S.~A.}\ \bibnamefont {Khrapak}},
  \bibinfo {author} {\bibfnamefont {H.~M.}\ \bibnamefont {Thomas}}, \ and\
  \bibinfo {author} {\bibfnamefont {G.~E.}\ \bibnamefont {Morfill}},\ }\href
  {\doibase 10.1039/c0sm00813c} {\bibfield  {journal} {\bibinfo  {journal}
  {Soft Matter}\ }\textbf {\bibinfo {volume} {7}},\ \bibinfo {pages} {1287}
  (\bibinfo {year} {2011})}\BibitemShut {NoStop}%
\bibitem [{\citenamefont {Nunomura}, \citenamefont {Samsonov},\ and\
  \citenamefont {Goree}(2000)}]{NunomuraPRL2000}%
  \BibitemOpen
  \bibfield  {author} {\bibinfo {author} {\bibfnamefont {S.}~\bibnamefont
  {Nunomura}}, \bibinfo {author} {\bibfnamefont {D.}~\bibnamefont {Samsonov}},
  \ and\ \bibinfo {author} {\bibfnamefont {J.}~\bibnamefont {Goree}},\ }\href
  {\doibase 10.1103/physrevlett.84.5141} {\bibfield  {journal} {\bibinfo
  {journal} {Phys. Rev. Lett.}\ }\textbf {\bibinfo {volume} {84}},\ \bibinfo
  {pages} {5141} (\bibinfo {year} {2000})}\BibitemShut {NoStop}%
\bibitem [{\citenamefont {Nunomura}\ \emph
  {et~al.}(2002{\natexlab{a}})\citenamefont {Nunomura}, \citenamefont {Goree},
  \citenamefont {Hu}, \citenamefont {Wang}, \citenamefont {Bhattacharjee},\
  and\ \citenamefont {Avinash}}]{NunomuraPRL2002}%
  \BibitemOpen
  \bibfield  {author} {\bibinfo {author} {\bibfnamefont {S.}~\bibnamefont
  {Nunomura}}, \bibinfo {author} {\bibfnamefont {J.}~\bibnamefont {Goree}},
  \bibinfo {author} {\bibfnamefont {S.}~\bibnamefont {Hu}}, \bibinfo {author}
  {\bibfnamefont {X.}~\bibnamefont {Wang}}, \bibinfo {author} {\bibfnamefont
  {A.}~\bibnamefont {Bhattacharjee}}, \ and\ \bibinfo {author} {\bibfnamefont
  {K.}~\bibnamefont {Avinash}},\ }\href {\doibase
  10.1103/physrevlett.89.035001} {\bibfield  {journal} {\bibinfo  {journal}
  {Phys. Rev. Lett.}\ }\textbf {\bibinfo {volume} {89}},\ \bibinfo {pages}
  {035001} (\bibinfo {year} {2002}{\natexlab{a}})}\BibitemShut {NoStop}%
\bibitem [{\citenamefont {Nunomura}\ \emph
  {et~al.}(2002{\natexlab{b}})\citenamefont {Nunomura}, \citenamefont {Goree},
  \citenamefont {Hu}, \citenamefont {Wang},\ and\ \citenamefont
  {Bhattacharjee}}]{NunomuraPRE2002}%
  \BibitemOpen
  \bibfield  {author} {\bibinfo {author} {\bibfnamefont {S.}~\bibnamefont
  {Nunomura}}, \bibinfo {author} {\bibfnamefont {J.}~\bibnamefont {Goree}},
  \bibinfo {author} {\bibfnamefont {S.}~\bibnamefont {Hu}}, \bibinfo {author}
  {\bibfnamefont {X.}~\bibnamefont {Wang}}, \ and\ \bibinfo {author}
  {\bibfnamefont {A.}~\bibnamefont {Bhattacharjee}},\ }\href {\doibase
  10.1103/physreve.65.066402} {\bibfield  {journal} {\bibinfo  {journal} {Phys.
  Rev. E}\ }\textbf {\bibinfo {volume} {65}},\ \bibinfo {pages} {066402}
  (\bibinfo {year} {2002}{\natexlab{b}})}\BibitemShut {NoStop}%
\bibitem [{\citenamefont {Peeters}\ and\ \citenamefont
  {Wu}(1987)}]{PeetersPRA1987}%
  \BibitemOpen
  \bibfield  {author} {\bibinfo {author} {\bibfnamefont {F.~M.}\ \bibnamefont
  {Peeters}}\ and\ \bibinfo {author} {\bibfnamefont {X.}~\bibnamefont {Wu}},\
  }\href {\doibase 10.1103/physreva.35.3109} {\bibfield  {journal} {\bibinfo
  {journal} {Phys. Rev. A}\ }\textbf {\bibinfo {volume} {35}},\ \bibinfo
  {pages} {3109} (\bibinfo {year} {1987})}\BibitemShut {NoStop}%
\bibitem [{\citenamefont {Dubin}(2000)}]{DubinPoP2000}%
  \BibitemOpen
  \bibfield  {author} {\bibinfo {author} {\bibfnamefont {D.~H.~E.}\
  \bibnamefont {Dubin}},\ }\href {\doibase 10.1063/1.1308078} {\bibfield
  {journal} {\bibinfo  {journal} {Phys. Plasmas}\ }\textbf {\bibinfo {volume}
  {7}},\ \bibinfo {pages} {3895} (\bibinfo {year} {2000})}\BibitemShut
  {NoStop}%
\bibitem [{\citenamefont {Wang}, \citenamefont {Bhattacharjee},\ and\
  \citenamefont {Hu}(2001)}]{WangPRL2001}%
  \BibitemOpen
  \bibfield  {author} {\bibinfo {author} {\bibfnamefont {X.}~\bibnamefont
  {Wang}}, \bibinfo {author} {\bibfnamefont {A.}~\bibnamefont {Bhattacharjee}},
  \ and\ \bibinfo {author} {\bibfnamefont {S.}~\bibnamefont {Hu}},\ }\href
  {\doibase 10.1103/physrevlett.86.2569} {\bibfield  {journal} {\bibinfo
  {journal} {Phys. Rev. Lett.}\ }\textbf {\bibinfo {volume} {86}},\ \bibinfo
  {pages} {2569} (\bibinfo {year} {2001})}\BibitemShut {NoStop}%
\bibitem [{\citenamefont {Totsuji}\ \emph {et~al.}(2004)\citenamefont
  {Totsuji}, \citenamefont {Liman}, \citenamefont {Totsuji},\ and\
  \citenamefont {Tsuruta}}]{TotsujiPRE2004}%
  \BibitemOpen
  \bibfield  {author} {\bibinfo {author} {\bibfnamefont {H.}~\bibnamefont
  {Totsuji}}, \bibinfo {author} {\bibfnamefont {M.~S.}\ \bibnamefont {Liman}},
  \bibinfo {author} {\bibfnamefont {C.}~\bibnamefont {Totsuji}}, \ and\
  \bibinfo {author} {\bibfnamefont {K.}~\bibnamefont {Tsuruta}},\ }\href
  {\doibase 10.1103/physreve.70.016405} {\bibfield  {journal} {\bibinfo
  {journal} {Phys. Rev. E}\ }\textbf {\bibinfo {volume} {70}},\ \bibinfo
  {pages} {016405} (\bibinfo {year} {2004})}\BibitemShut {NoStop}%
\bibitem [{\citenamefont {Hartmann}\ \emph {et~al.}(2005)\citenamefont
  {Hartmann}, \citenamefont {Kalman}, \citenamefont {Donk{\'{o}}},\ and\
  \citenamefont {Kutasi}}]{HartmannPRE2005}%
  \BibitemOpen
  \bibfield  {author} {\bibinfo {author} {\bibfnamefont {P.}~\bibnamefont
  {Hartmann}}, \bibinfo {author} {\bibfnamefont {G.~J.}\ \bibnamefont
  {Kalman}}, \bibinfo {author} {\bibfnamefont {Z.}~\bibnamefont {Donk{\'{o}}}},
  \ and\ \bibinfo {author} {\bibfnamefont {K.}~\bibnamefont {Kutasi}},\ }\href
  {\doibase 10.1103/physreve.72.026409} {\bibfield  {journal} {\bibinfo
  {journal} {Phys. Rev. E}\ }\textbf {\bibinfo {volume} {72}},\ \bibinfo
  {pages} {026409} (\bibinfo {year} {2005})}\BibitemShut {NoStop}%
\bibitem [{\citenamefont {Vaulina}\ \emph {et~al.}(2010)\citenamefont
  {Vaulina}, \citenamefont {Koss}, \citenamefont {Khrustalyov}, \citenamefont
  {Petrov},\ and\ \citenamefont {Fortov}}]{VaulinaPRE2010}%
  \BibitemOpen
  \bibfield  {author} {\bibinfo {author} {\bibfnamefont {O.~S.}\ \bibnamefont
  {Vaulina}}, \bibinfo {author} {\bibfnamefont {X.~G.}\ \bibnamefont {Koss}},
  \bibinfo {author} {\bibfnamefont {Y.~V.}\ \bibnamefont {Khrustalyov}},
  \bibinfo {author} {\bibfnamefont {O.~F.}\ \bibnamefont {Petrov}}, \ and\
  \bibinfo {author} {\bibfnamefont {V.~E.}\ \bibnamefont {Fortov}},\ }\href
  {\doibase 10.1103/physreve.82.056411} {\bibfield  {journal} {\bibinfo
  {journal} {Phys. Rev. E}\ }\textbf {\bibinfo {volume} {82}},\ \bibinfo
  {pages} {056411} (\bibinfo {year} {2010})}\BibitemShut {NoStop}%
\bibitem [{\citenamefont {Semenov}, \citenamefont {Khrapak},\ and\
  \citenamefont {Thomas}(2015)}]{SemenovPoP2015}%
  \BibitemOpen
  \bibfield  {author} {\bibinfo {author} {\bibfnamefont {I.~L.}\ \bibnamefont
  {Semenov}}, \bibinfo {author} {\bibfnamefont {S.~A.}\ \bibnamefont
  {Khrapak}}, \ and\ \bibinfo {author} {\bibfnamefont {H.~M.}\ \bibnamefont
  {Thomas}},\ }\href {\doibase 10.1063/1.4935846} {\bibfield  {journal}
  {\bibinfo  {journal} {Phys. Plasmas}\ }\textbf {\bibinfo {volume} {22}},\
  \bibinfo {pages} {114504} (\bibinfo {year} {2015})}\BibitemShut {NoStop}%
\bibitem [{\citenamefont {Khrapak}\ \emph {et~al.}(2015)\citenamefont
  {Khrapak}, \citenamefont {Semenov}, \citenamefont {Couëdel},\ and\
  \citenamefont {Thomas}}]{KhrapakPoP08_2015}%
  \BibitemOpen
  \bibfield  {author} {\bibinfo {author} {\bibfnamefont {S.~A.}\ \bibnamefont
  {Khrapak}}, \bibinfo {author} {\bibfnamefont {I.~L.}\ \bibnamefont
  {Semenov}}, \bibinfo {author} {\bibfnamefont {L.}~\bibnamefont {Couëdel}}, \
  and\ \bibinfo {author} {\bibfnamefont {H.~M.}\ \bibnamefont {Thomas}},\
  }\href {\doibase 10.1063/1.4928443} {\bibfield  {journal} {\bibinfo
  {journal} {Phys. Plasmas}\ }\textbf {\bibinfo {volume} {22}},\ \bibinfo
  {pages} {083706} (\bibinfo {year} {2015})}\BibitemShut {NoStop}%
\bibitem [{\citenamefont {Feng}\ \emph {et~al.}(2016)\citenamefont {Feng},
  \citenamefont {Goree}, \citenamefont {Liu}, \citenamefont {Wang},\ and\
  \citenamefont {de~Tian}}]{FengJPD2016}%
  \BibitemOpen
  \bibfield  {author} {\bibinfo {author} {\bibfnamefont {Y.}~\bibnamefont
  {Feng}}, \bibinfo {author} {\bibfnamefont {J.}~\bibnamefont {Goree}},
  \bibinfo {author} {\bibfnamefont {B.}~\bibnamefont {Liu}}, \bibinfo {author}
  {\bibfnamefont {L.}~\bibnamefont {Wang}}, \ and\ \bibinfo {author}
  {\bibfnamefont {W.}~\bibnamefont {de~Tian}},\ }\href {\doibase
  10.1088/0022-3727/49/23/235203} {\bibfield  {journal} {\bibinfo  {journal}
  {Journal of Physics D: Applied Physics}\ }\textbf {\bibinfo {volume} {49}},\
  \bibinfo {pages} {235203} (\bibinfo {year} {2016})}\BibitemShut {NoStop}%
\bibitem [{\citenamefont {Yurchenko}, \citenamefont {Kryuchkov},\ and\
  \citenamefont {Ivlev}(2015)}]{YurchenkoJCP2015}%
  \BibitemOpen
  \bibfield  {author} {\bibinfo {author} {\bibfnamefont {S.~O.}\ \bibnamefont
  {Yurchenko}}, \bibinfo {author} {\bibfnamefont {N.~P.}\ \bibnamefont
  {Kryuchkov}}, \ and\ \bibinfo {author} {\bibfnamefont {A.~V.}\ \bibnamefont
  {Ivlev}},\ }\href {\doibase 10.1063/1.4926945} {\bibfield  {journal}
  {\bibinfo  {journal} {J. Chem. Phys.}\ }\textbf {\bibinfo {volume} {143}},\
  \bibinfo {pages} {034506} (\bibinfo {year} {2015})}\BibitemShut {NoStop}%
\bibitem [{\citenamefont {Yurchenko}, \citenamefont {Kryuchkov},\ and\
  \citenamefont {Ivlev}(2016)}]{YurchenkoJPCM2016}%
  \BibitemOpen
  \bibfield  {author} {\bibinfo {author} {\bibfnamefont {S.~O.}\ \bibnamefont
  {Yurchenko}}, \bibinfo {author} {\bibfnamefont {N.~P.}\ \bibnamefont
  {Kryuchkov}}, \ and\ \bibinfo {author} {\bibfnamefont {A.~V.}\ \bibnamefont
  {Ivlev}},\ }\href {\doibase 10.1088/0953-8984/28/23/235401} {\bibfield
  {journal} {\bibinfo  {journal} {J. Phys.: Condens. Matter}\ }\textbf
  {\bibinfo {volume} {28}},\ \bibinfo {pages} {235401} (\bibinfo {year}
  {2016})}\BibitemShut {NoStop}%
\bibitem [{\citenamefont {Kryuchkov}, \citenamefont {Khrapak},\ and\
  \citenamefont {Yurchenko}(2017)}]{KryuchkovJCP2017}%
  \BibitemOpen
  \bibfield  {author} {\bibinfo {author} {\bibfnamefont {N.~P.}\ \bibnamefont
  {Kryuchkov}}, \bibinfo {author} {\bibfnamefont {S.~A.}\ \bibnamefont
  {Khrapak}}, \ and\ \bibinfo {author} {\bibfnamefont {S.~O.}\ \bibnamefont
  {Yurchenko}},\ }\href {\doibase 10.1063/1.4979325} {\bibfield  {journal}
  {\bibinfo  {journal} {J. Chem. Phys.}\ }\textbf {\bibinfo {volume} {146}},\
  \bibinfo {pages} {134702} (\bibinfo {year} {2017})}\BibitemShut {NoStop}%
\bibitem [{\citenamefont {Caillol}\ \emph {et~al.}(1982)\citenamefont
  {Caillol}, \citenamefont {Levesque}, \citenamefont {Weis},\ and\
  \citenamefont {Hansen}}]{Caillol1982}%
  \BibitemOpen
  \bibfield  {author} {\bibinfo {author} {\bibfnamefont {J.~M.}\ \bibnamefont
  {Caillol}}, \bibinfo {author} {\bibfnamefont {D.}~\bibnamefont {Levesque}},
  \bibinfo {author} {\bibfnamefont {J.~J.}\ \bibnamefont {Weis}}, \ and\
  \bibinfo {author} {\bibfnamefont {J.~P.}\ \bibnamefont {Hansen}},\ }\href
  {\doibase 10.1007/bf01012609} {\bibfield  {journal} {\bibinfo  {journal} {J.
  Stat. Phys.}\ }\textbf {\bibinfo {volume} {28}},\ \bibinfo {pages} {325}
  (\bibinfo {year} {1982})}\BibitemShut {NoStop}%
\bibitem [{\citenamefont {Khrapak}\ and\ \citenamefont
  {Khrapak}(2016)}]{KhrapakCPP2016}%
  \BibitemOpen
  \bibfield  {author} {\bibinfo {author} {\bibfnamefont {S.~A.}\ \bibnamefont
  {Khrapak}}\ and\ \bibinfo {author} {\bibfnamefont {A.~G.}\ \bibnamefont
  {Khrapak}},\ }\href {\doibase 10.1002/ctpp.201500104} {\bibfield  {journal}
  {\bibinfo  {journal} {Contrib. Plasma Phys.}\ }\textbf {\bibinfo {volume}
  {56}},\ \bibinfo {pages} {270} (\bibinfo {year} {2016})}\BibitemShut
  {NoStop}%
\bibitem [{\citenamefont {Gann}, \citenamefont {Chakravarty},\ and\
  \citenamefont {Chester}(1979)}]{GannPRB1979}%
  \BibitemOpen
  \bibfield  {author} {\bibinfo {author} {\bibfnamefont {R.~C.}\ \bibnamefont
  {Gann}}, \bibinfo {author} {\bibfnamefont {S.}~\bibnamefont {Chakravarty}}, \
  and\ \bibinfo {author} {\bibfnamefont {G.~V.}\ \bibnamefont {Chester}},\
  }\href {\doibase 10.1103/physrevb.20.326} {\bibfield  {journal} {\bibinfo
  {journal} {Phys. Rev. B}\ }\textbf {\bibinfo {volume} {20}},\ \bibinfo
  {pages} {326} (\bibinfo {year} {1979})}\BibitemShut {NoStop}%
\bibitem [{\citenamefont {Grimes}\ and\ \citenamefont
  {Adams}(1979)}]{GrimesPRL1979}%
  \BibitemOpen
  \bibfield  {author} {\bibinfo {author} {\bibfnamefont {C.~C.}\ \bibnamefont
  {Grimes}}\ and\ \bibinfo {author} {\bibfnamefont {G.}~\bibnamefont {Adams}},\
  }\href {\doibase 10.1103/physrevlett.42.795} {\bibfield  {journal} {\bibinfo
  {journal} {Phys. Rev. Lett.}\ }\textbf {\bibinfo {volume} {42}},\ \bibinfo
  {pages} {795} (\bibinfo {year} {1979})}\BibitemShut {NoStop}%
\bibitem [{\citenamefont {Kosterlitz}(2017)}]{KosterlitzRMP2017}%
  \BibitemOpen
  \bibfield  {author} {\bibinfo {author} {\bibfnamefont {J.~M.}\ \bibnamefont
  {Kosterlitz}},\ }\href {\doibase 10.1103/revmodphys.89.040501} {\bibfield
  {journal} {\bibinfo  {journal} {Rev. Mod. Phys.}\ }\textbf {\bibinfo {volume}
  {89}},\ \bibinfo {pages} {040501} (\bibinfo {year} {2017})}\BibitemShut
  {NoStop}%
\bibitem [{\citenamefont {Zahn}, \citenamefont {Lenke},\ and\ \citenamefont
  {Maret}(1999)}]{ZahnPRL1999}%
  \BibitemOpen
  \bibfield  {author} {\bibinfo {author} {\bibfnamefont {K.}~\bibnamefont
  {Zahn}}, \bibinfo {author} {\bibfnamefont {R.}~\bibnamefont {Lenke}}, \ and\
  \bibinfo {author} {\bibfnamefont {G.}~\bibnamefont {Maret}},\ }\href
  {\doibase 10.1103/physrevlett.82.2721} {\bibfield  {journal} {\bibinfo
  {journal} {Phys. Rev. Lett.}\ }\textbf {\bibinfo {volume} {82}},\ \bibinfo
  {pages} {2721} (\bibinfo {year} {1999})}\BibitemShut {NoStop}%
\bibitem [{\citenamefont {von Gr\"{u}nberg}\ \emph {et~al.}(2004)\citenamefont
  {von Gr\"{u}nberg}, \citenamefont {Keim}, \citenamefont {Zahn},\ and\
  \citenamefont {Maret}}]{GrunbergPRL2004}%
  \BibitemOpen
  \bibfield  {author} {\bibinfo {author} {\bibfnamefont {H.~H.}\ \bibnamefont
  {von Gr\"{u}nberg}}, \bibinfo {author} {\bibfnamefont {P.}~\bibnamefont
  {Keim}}, \bibinfo {author} {\bibfnamefont {K.}~\bibnamefont {Zahn}}, \ and\
  \bibinfo {author} {\bibfnamefont {G.}~\bibnamefont {Maret}},\ }\href
  {\doibase 10.1103/physrevlett.93.255703} {\bibfield  {journal} {\bibinfo
  {journal} {Phys. Rev. Lett.}\ }\textbf {\bibinfo {volume} {93}},\ \bibinfo
  {pages} {255703} (\bibinfo {year} {2004})}\BibitemShut {NoStop}%
\bibitem [{\citenamefont {Zanghellini}, \citenamefont {Keim},\ and\
  \citenamefont {von Gr\"{u}nberg}(2005)}]{ZanghelliniJPCM2005}%
  \BibitemOpen
  \bibfield  {author} {\bibinfo {author} {\bibfnamefont {J.}~\bibnamefont
  {Zanghellini}}, \bibinfo {author} {\bibfnamefont {P.}~\bibnamefont {Keim}}, \
  and\ \bibinfo {author} {\bibfnamefont {H.~H.}\ \bibnamefont {von
  Gr\"{u}nberg}},\ }\href {\doibase 10.1088/0953-8984/17/45/051} {\bibfield
  {journal} {\bibinfo  {journal} {J. Phys.: Condens. Matter}\ }\textbf
  {\bibinfo {volume} {17}},\ \bibinfo {pages} {S3579} (\bibinfo {year}
  {2005})}\BibitemShut {NoStop}%
\bibitem [{\citenamefont {Kapfer}\ and\ \citenamefont
  {Krauth}(2015)}]{KapferPRL2015}%
  \BibitemOpen
  \bibfield  {author} {\bibinfo {author} {\bibfnamefont {S.~C.}\ \bibnamefont
  {Kapfer}}\ and\ \bibinfo {author} {\bibfnamefont {W.}~\bibnamefont
  {Krauth}},\ }\href {\doibase 10.1103/physrevlett.114.035702} {\bibfield
  {journal} {\bibinfo  {journal} {Phys. Rev. Lett.}\ }\textbf {\bibinfo
  {volume} {114}},\ \bibinfo {pages} {035702} (\bibinfo {year}
  {2015})}\BibitemShut {NoStop}%
\bibitem [{\citenamefont {Bernard}\ and\ \citenamefont
  {Krauth}(2011)}]{BernardPRL2011}%
  \BibitemOpen
  \bibfield  {author} {\bibinfo {author} {\bibfnamefont {E.~P.}\ \bibnamefont
  {Bernard}}\ and\ \bibinfo {author} {\bibfnamefont {W.}~\bibnamefont
  {Krauth}},\ }\href {\doibase 10.1103/physrevlett.107.155704} {\bibfield
  {journal} {\bibinfo  {journal} {Phys. Rev. Lett.}\ }\textbf {\bibinfo
  {volume} {107}},\ \bibinfo {pages} {155704} (\bibinfo {year}
  {2011})}\BibitemShut {NoStop}%
\bibitem [{\citenamefont {Engel}\ \emph {et~al.}(2013)\citenamefont {Engel},
  \citenamefont {Anderson}, \citenamefont {Glotzer}, \citenamefont {Isobe},
  \citenamefont {Bernard},\ and\ \citenamefont {Krauth}}]{EngelPRE2013}%
  \BibitemOpen
  \bibfield  {author} {\bibinfo {author} {\bibfnamefont {M.}~\bibnamefont
  {Engel}}, \bibinfo {author} {\bibfnamefont {J.~A.}\ \bibnamefont {Anderson}},
  \bibinfo {author} {\bibfnamefont {S.~C.}\ \bibnamefont {Glotzer}}, \bibinfo
  {author} {\bibfnamefont {M.}~\bibnamefont {Isobe}}, \bibinfo {author}
  {\bibfnamefont {E.~P.}\ \bibnamefont {Bernard}}, \ and\ \bibinfo {author}
  {\bibfnamefont {W.}~\bibnamefont {Krauth}},\ }\href {\doibase
  10.1103/physreve.87.042134} {\bibfield  {journal} {\bibinfo  {journal} {Phys.
  Rev. E}\ }\textbf {\bibinfo {volume} {87}},\ \bibinfo {pages} {042134}
  (\bibinfo {year} {2013})}\BibitemShut {NoStop}%
\bibitem [{\citenamefont {Thorneywork}\ \emph {et~al.}(2017)\citenamefont
  {Thorneywork}, \citenamefont {Abbott}, \citenamefont {Aarts},\ and\
  \citenamefont {Dullens}}]{ThorneyworkPRL2017}%
  \BibitemOpen
  \bibfield  {author} {\bibinfo {author} {\bibfnamefont {A.~L.}\ \bibnamefont
  {Thorneywork}}, \bibinfo {author} {\bibfnamefont {J.~L.}\ \bibnamefont
  {Abbott}}, \bibinfo {author} {\bibfnamefont {D.~G.}\ \bibnamefont {Aarts}}, \
  and\ \bibinfo {author} {\bibfnamefont {R.~P.}\ \bibnamefont {Dullens}},\
  }\href {\doibase 10.1103/physrevlett.118.158001} {\bibfield  {journal}
  {\bibinfo  {journal} {Phys. Rev. Lett.}\ }\textbf {\bibinfo {volume} {118}},\
  \bibinfo {pages} {158001} (\bibinfo {year} {2017})}\BibitemShut {NoStop}%
\bibitem [{\citenamefont {Nosenko}\ and\ \citenamefont
  {Goree}(2004)}]{NosenkoPRL2004}%
  \BibitemOpen
  \bibfield  {author} {\bibinfo {author} {\bibfnamefont {V.}~\bibnamefont
  {Nosenko}}\ and\ \bibinfo {author} {\bibfnamefont {J.}~\bibnamefont
  {Goree}},\ }\href {\doibase 10.1103/physrevlett.93.155004} {\bibfield
  {journal} {\bibinfo  {journal} {Phys. Rev. Lett.}\ }\textbf {\bibinfo
  {volume} {93}},\ \bibinfo {pages} {155004} (\bibinfo {year}
  {2004})}\BibitemShut {NoStop}%
\bibitem [{\citenamefont {Nosenko}\ \emph {et~al.}(2009)\citenamefont
  {Nosenko}, \citenamefont {Zhdanov}, \citenamefont {Ivlev}, \citenamefont
  {Knapek},\ and\ \citenamefont {Morfill}}]{NosenkoPRL2009}%
  \BibitemOpen
  \bibfield  {author} {\bibinfo {author} {\bibfnamefont {V.}~\bibnamefont
  {Nosenko}}, \bibinfo {author} {\bibfnamefont {S.~K.}\ \bibnamefont
  {Zhdanov}}, \bibinfo {author} {\bibfnamefont {A.~V.}\ \bibnamefont {Ivlev}},
  \bibinfo {author} {\bibfnamefont {C.~A.}\ \bibnamefont {Knapek}}, \ and\
  \bibinfo {author} {\bibfnamefont {G.~E.}\ \bibnamefont {Morfill}},\ }\href
  {\doibase 10.1103/physrevlett.103.015001} {\bibfield  {journal} {\bibinfo
  {journal} {Phys. Rev. Lett.}\ }\textbf {\bibinfo {volume} {103}},\ \bibinfo
  {pages} {015001} (\bibinfo {year} {2009})}\BibitemShut {NoStop}%
\bibitem [{\citenamefont {Zwanzig}\ and\ \citenamefont
  {Mountain}(1965)}]{ZwanzigJCP1965}%
  \BibitemOpen
  \bibfield  {author} {\bibinfo {author} {\bibfnamefont {R.}~\bibnamefont
  {Zwanzig}}\ and\ \bibinfo {author} {\bibfnamefont {R.~D.}\ \bibnamefont
  {Mountain}},\ }\href {\doibase 10.1063/1.1696718} {\bibfield  {journal}
  {\bibinfo  {journal} {J. Chem. Phys.}\ }\textbf {\bibinfo {volume} {43}},\
  \bibinfo {pages} {4464} (\bibinfo {year} {1965})}\BibitemShut {NoStop}%
\bibitem [{\citenamefont {Khrapak}, \citenamefont {Kryuchkov},\ and\
  \citenamefont {Yurchenko}(2018)}]{IPL3}%
  \BibitemOpen
  \bibfield  {author} {\bibinfo {author} {\bibfnamefont {S.}~\bibnamefont
  {Khrapak}}, \bibinfo {author} {\bibfnamefont {N.}~\bibnamefont {Kryuchkov}},
  \ and\ \bibinfo {author} {\bibfnamefont {S.}~\bibnamefont {Yurchenko}},\
  }\href@noop {} {\bibfield  {journal} {\bibinfo  {journal} {Phys. Rev. E}\
  }\textbf {\bibinfo {volume} {97}},\ \bibinfo {pages} {022616} (\bibinfo
  {year} {2018})}\BibitemShut {NoStop}%
\bibitem [{\citenamefont {Khrapak}(2016{\natexlab{a}})}]{KhrapakOnset}%
  \BibitemOpen
  \bibfield  {author} {\bibinfo {author} {\bibfnamefont {S.~A.}\ \bibnamefont
  {Khrapak}},\ }\href {\doibase 10.1063/1.4965903} {\bibfield  {journal}
  {\bibinfo  {journal} {Phys. Plasmas}\ }\textbf {\bibinfo {volume} {23}},\
  \bibinfo {pages} {104506} (\bibinfo {year} {2016}{\natexlab{a}})}\BibitemShut
  {NoStop}%
\bibitem [{\citenamefont {Khrapak}, \citenamefont {Klumov},\ and\ \citenamefont
  {Couedel}(2017)}]{KhrapakSciRep2017}%
  \BibitemOpen
  \bibfield  {author} {\bibinfo {author} {\bibfnamefont {S.}~\bibnamefont
  {Khrapak}}, \bibinfo {author} {\bibfnamefont {B.}~\bibnamefont {Klumov}}, \
  and\ \bibinfo {author} {\bibfnamefont {L.}~\bibnamefont {Couedel}},\
  }\href@noop {} {\bibfield  {journal} {\bibinfo  {journal} {Sci. Reports}\
  }\textbf {\bibinfo {volume} {7}},\ \bibinfo {pages} {7985} (\bibinfo {year}
  {2017})}\BibitemShut {NoStop}%
\bibitem [{\citenamefont {Hansen}\ and\ \citenamefont
  {MacDonald}(2006)}]{Hansen_Book}%
  \BibitemOpen
  \bibfield  {author} {\bibinfo {author} {\bibfnamefont {J.-P.}\ \bibnamefont
  {Hansen}}\ and\ \bibinfo {author} {\bibfnamefont {I.~R.}\ \bibnamefont
  {MacDonald}},\ }\href@noop {} {\emph {\bibinfo {title} {Theory of simple
  liquids}}}\ (\bibinfo  {publisher} {London: Academic},\ \bibinfo {year}
  {2006})\BibitemShut {NoStop}%
\bibitem [{\citenamefont {Frenkel}\ and\ \citenamefont
  {Smit}(2001)}]{Frenkel2001}%
  \BibitemOpen
  \bibfield  {author} {\bibinfo {author} {\bibfnamefont {D.}~\bibnamefont
  {Frenkel}}\ and\ \bibinfo {author} {\bibfnamefont {B.}~\bibnamefont {Smit}},\
  }\href@noop {} {\emph {\bibinfo {title} {Understanding Molecular Simulation:
  From Algorithms to Applications}}}\ (\bibinfo  {publisher} {Elsevier
  Science},\ \bibinfo {year} {2001})\BibitemShut {NoStop}%
\bibitem [{\citenamefont {Khrapak}(2016{\natexlab{b}})}]{QCA_Relations}%
  \BibitemOpen
  \bibfield  {author} {\bibinfo {author} {\bibfnamefont {S.~A.}\ \bibnamefont
  {Khrapak}},\ }\href {\doibase 10.1063/1.4942171} {\bibfield  {journal}
  {\bibinfo  {journal} {Phys. Plasmas}\ }\textbf {\bibinfo {volume} {23}},\
  \bibinfo {pages} {024504} (\bibinfo {year} {2016}{\natexlab{b}})}\BibitemShut
  {NoStop}%
\bibitem [{\citenamefont {Schofield}(1966)}]{Schofield1966}%
  \BibitemOpen
  \bibfield  {author} {\bibinfo {author} {\bibfnamefont {P.}~\bibnamefont
  {Schofield}},\ }\href {\doibase 10.1088/0370-1328/88/1/318} {\bibfield
  {journal} {\bibinfo  {journal} {Proc. Phys. Soc.}\ }\textbf {\bibinfo
  {volume} {88}},\ \bibinfo {pages} {149} (\bibinfo {year} {1966})}\BibitemShut
  {NoStop}%
\bibitem [{\citenamefont {Landau}\ and\ \citenamefont
  {Lifshitz}(2013)}]{LL_Hydrodynamics}%
  \BibitemOpen
  \bibfield  {author} {\bibinfo {author} {\bibfnamefont {L.~D.}\ \bibnamefont
  {Landau}}\ and\ \bibinfo {author} {\bibfnamefont {E.~M.}\ \bibnamefont
  {Lifshitz}},\ }\href@noop {} {\emph {\bibinfo {title} {Fluid Mechanics:
  Volume 6 (Pergamon International Library of Science, Technology, Engineering
  \& Social Studies)}}}\ (\bibinfo  {publisher} {Pergamon},\ \bibinfo {year}
  {2013})\BibitemShut {NoStop}%
\bibitem [{\citenamefont {Khrapak}, \citenamefont {Klumov},\ and\ \citenamefont
  {Khrapak}(2016)}]{Khrapak2016_2DOCP}%
  \BibitemOpen
  \bibfield  {author} {\bibinfo {author} {\bibfnamefont {S.~A.}\ \bibnamefont
  {Khrapak}}, \bibinfo {author} {\bibfnamefont {B.~A.}\ \bibnamefont {Klumov}},
  \ and\ \bibinfo {author} {\bibfnamefont {A.~G.}\ \bibnamefont {Khrapak}},\
  }\href {\doibase 10.1063/1.4950829} {\bibfield  {journal} {\bibinfo
  {journal} {Phys. Plasmas}\ }\textbf {\bibinfo {volume} {23}},\ \bibinfo
  {pages} {052115} (\bibinfo {year} {2016})}\BibitemShut {NoStop}%
\bibitem [{\citenamefont {Khrapak}\ and\ \citenamefont
  {Thomas}(2015)}]{KhrapakPRE2015_Sound}%
  \BibitemOpen
  \bibfield  {author} {\bibinfo {author} {\bibfnamefont {S.~A.}\ \bibnamefont
  {Khrapak}}\ and\ \bibinfo {author} {\bibfnamefont {H.~M.}\ \bibnamefont
  {Thomas}},\ }\href@noop {} {\bibfield  {journal} {\bibinfo  {journal} {Phys.
  Rev. E}\ }\textbf {\bibinfo {volume} {91}},\ \bibinfo {pages} {033110}
  (\bibinfo {year} {2015})}\BibitemShut {NoStop}%
\bibitem [{\citenamefont {Khrapak}(2016{\natexlab{c}})}]{KhrapakPPCF2016}%
  \BibitemOpen
  \bibfield  {author} {\bibinfo {author} {\bibfnamefont {S.~A.}\ \bibnamefont
  {Khrapak}},\ }\href@noop {} {\bibfield  {journal} {\bibinfo  {journal}
  {Plasma Phys. Controlled Fusion}\ }\textbf {\bibinfo {volume} {58}},\
  \bibinfo {pages} {014022} (\bibinfo {year} {2016}{\natexlab{c}})}\BibitemShut
  {NoStop}%
\bibitem [{\citenamefont {Khrapak}(2017)}]{KhrapakAIPAdv2017}%
  \BibitemOpen
  \bibfield  {author} {\bibinfo {author} {\bibfnamefont {S.~A.}\ \bibnamefont
  {Khrapak}},\ }\href {\doibase 10.1063/1.5002130} {\bibfield  {journal}
  {\bibinfo  {journal} {{AIP} Adv.}\ }\textbf {\bibinfo {volume} {7}},\
  \bibinfo {pages} {125026} (\bibinfo {year} {2017})}\BibitemShut {NoStop}%
\bibitem [{\citenamefont {Veldhorst}, \citenamefont {Schr{\o}der},\ and\
  \citenamefont {Dyre}(2015)}]{VeldhorstPoP2015}%
  \BibitemOpen
  \bibfield  {author} {\bibinfo {author} {\bibfnamefont {A.~A.}\ \bibnamefont
  {Veldhorst}}, \bibinfo {author} {\bibfnamefont {T.~B.}\ \bibnamefont
  {Schr{\o}der}}, \ and\ \bibinfo {author} {\bibfnamefont {J.~C.}\ \bibnamefont
  {Dyre}},\ }\href {\doibase 10.1063/1.4926822} {\bibfield  {journal} {\bibinfo
   {journal} {Phys. Plasmas}\ }\textbf {\bibinfo {volume} {22}},\ \bibinfo
  {pages} {073705} (\bibinfo {year} {2015})}\BibitemShut {NoStop}%
\bibitem [{\citenamefont {Farouki}\ and\ \citenamefont
  {Hamaguchi}(1994)}]{FaroukiJCP1994}%
  \BibitemOpen
  \bibfield  {author} {\bibinfo {author} {\bibfnamefont {R.~T.}\ \bibnamefont
  {Farouki}}\ and\ \bibinfo {author} {\bibfnamefont {S.}~\bibnamefont
  {Hamaguchi}},\ }\href {\doibase 10.1063/1.467955} {\bibfield  {journal}
  {\bibinfo  {journal} {J. Chem. Phys.}\ }\textbf {\bibinfo {volume} {101}},\
  \bibinfo {pages} {9885} (\bibinfo {year} {1994})}\BibitemShut {NoStop}%
\bibitem [{\citenamefont {Pereira}\ and\ \citenamefont
  {Apolinario}(2012)}]{PereiraPRE2012}%
  \BibitemOpen
  \bibfield  {author} {\bibinfo {author} {\bibfnamefont {P.~C.~N.}\
  \bibnamefont {Pereira}}\ and\ \bibinfo {author} {\bibfnamefont {S.~W.~S.}\
  \bibnamefont {Apolinario}},\ }\href {\doibase 10.1103/physreve.86.046702}
  {\bibfield  {journal} {\bibinfo  {journal} {Phys. Rev. E}\ }\textbf {\bibinfo
  {volume} {86}},\ \bibinfo {pages} {046702} (\bibinfo {year}
  {2012})}\BibitemShut {NoStop}%
\bibitem [{\citenamefont {Baus}\ and\ \citenamefont
  {Hansen}(1980)}]{BausPR1980}%
  \BibitemOpen
  \bibfield  {author} {\bibinfo {author} {\bibfnamefont {M.}~\bibnamefont
  {Baus}}\ and\ \bibinfo {author} {\bibfnamefont {J.~P.}\ \bibnamefont
  {Hansen}},\ }\href {\doibase 10.1016/0370-1573(80)90022-8} {\bibfield
  {journal} {\bibinfo  {journal} {Phys. Rep.}\ }\textbf {\bibinfo {volume}
  {59}},\ \bibinfo {pages} {1} (\bibinfo {year} {1980})}\BibitemShut {NoStop}%
\bibitem [{\citenamefont {Knapek}\ \emph {et~al.}(2007)\citenamefont {Knapek},
  \citenamefont {Ivlev}, \citenamefont {Klumov}, \citenamefont {Morfill},\ and\
  \citenamefont {Samsonov}}]{KnapekPRL2007}%
  \BibitemOpen
  \bibfield  {author} {\bibinfo {author} {\bibfnamefont {C.~A.}\ \bibnamefont
  {Knapek}}, \bibinfo {author} {\bibfnamefont {A.~V.}\ \bibnamefont {Ivlev}},
  \bibinfo {author} {\bibfnamefont {B.~A.}\ \bibnamefont {Klumov}}, \bibinfo
  {author} {\bibfnamefont {G.~E.}\ \bibnamefont {Morfill}}, \ and\ \bibinfo
  {author} {\bibfnamefont {D.}~\bibnamefont {Samsonov}},\ }\href {\doibase
  10.1103/physrevlett.98.015001} {\bibfield  {journal} {\bibinfo  {journal}
  {Phys. Rev. Lett.}\ }\textbf {\bibinfo {volume} {98}},\ \bibinfo {pages}
  {015001} (\bibinfo {year} {2007})}\BibitemShut {NoStop}%
\bibitem [{\citenamefont {Wong}, \citenamefont {Goree},\ and\ \citenamefont
  {Haralson}(2017)}]{WongIEEE2017}%
  \BibitemOpen
  \bibfield  {author} {\bibinfo {author} {\bibfnamefont {C.-S.}\ \bibnamefont
  {Wong}}, \bibinfo {author} {\bibfnamefont {J.}~\bibnamefont {Goree}}, \ and\
  \bibinfo {author} {\bibfnamefont {Z.}~\bibnamefont {Haralson}},\ }\href
  {\doibase 10.1109/tps.2017.2746012} {\bibfield  {journal} {\bibinfo
  {journal} {{IEEE} Trans. on Plasma Sci.}\ }\textbf {\bibinfo {volume} {XX}},\
  \bibinfo {pages} {xxxx} (\bibinfo {year} {2017})}\BibitemShut {NoStop}%
\bibitem [{\citenamefont {Topping}(1927)}]{Topping}%
  \BibitemOpen
  \bibfield  {author} {\bibinfo {author} {\bibfnamefont {J.}~\bibnamefont
  {Topping}},\ }\href {http://www.jstor.org/stable/94698} {\bibfield  {journal}
  {\bibinfo  {journal} {Proceedings of the Royal Society of London}\ }\textbf
  {\bibinfo {volume} {A114}},\ \bibinfo {pages} {67} (\bibinfo {year}
  {1927})}\BibitemShut {NoStop}%
\bibitem [{\citenamefont {van~der Hoff}\ and\ \citenamefont
  {Benson}(1953)}]{derHoff}%
  \BibitemOpen
  \bibfield  {author} {\bibinfo {author} {\bibfnamefont {B.~M.~E.}\
  \bibnamefont {van~der Hoff}}\ and\ \bibinfo {author} {\bibfnamefont {G.~C.}\
  \bibnamefont {Benson}},\ }\href {\doibase 10.1139/p53-093} {\bibfield
  {journal} {\bibinfo  {journal} {Canadian J. Phys.}\ }\textbf {\bibinfo
  {volume} {31}},\ \bibinfo {pages} {1087} (\bibinfo {year}
  {1953})}\BibitemShut {NoStop}%
\bibitem [{\citenamefont {Donko}, \citenamefont {Kalman},\ and\ \citenamefont
  {Hartmann}(2008)}]{DonkoJPCM2008}%
  \BibitemOpen
  \bibfield  {author} {\bibinfo {author} {\bibfnamefont {Z.}~\bibnamefont
  {Donko}}, \bibinfo {author} {\bibfnamefont {G.~J.}\ \bibnamefont {Kalman}}, \
  and\ \bibinfo {author} {\bibfnamefont {P.}~\bibnamefont {Hartmann}},\
  }\href@noop {} {\bibfield  {journal} {\bibinfo  {journal} {J. Phys.: Condens.
  Matter}\ }\textbf {\bibinfo {volume} {20}},\ \bibinfo {pages} {413101}
  (\bibinfo {year} {2008})}\BibitemShut {NoStop}%
\end{thebibliography}%

\end{document}